\documentclass[%
superscriptaddress,
 amsmath,amssymb,
 aps,
prb,
twocolumn,
floatfix,
notitlepage
]{revtex4-1}

\usepackage{graphicx,placeins}
\usepackage{dcolumn}
\usepackage{bm}
\usepackage{subfigure}
\usepackage{units,ulem}
\usepackage[usenames, dvipsnames]{color}
\usepackage[colorlinks=true,bookmarks=true,%
     pdfborder={0 0 0},linkcolor=blue,urlcolor=red,%
     breaklinks=true,bookmarksnumbered=true]{hyperref}
     \hypersetup{
    colorlinks,
    citecolor=red,
    filecolor=cyan,
    linkcolor=blue,
    urlcolor=magenta
}

\def \beq{\begin{equation}}
\def \eeq{\end{equation}}
\def \beqarr{\begin{eqnarray}}
\def \eeqarr{\end{eqnarray}}
\def \bspt{\begin{split}}
\def \espt{\end{split}}
\def \bef{\begin{figure}}
\def \enf{\end{figure}}

\def \bpm{\begin{pmatrix}}
\def \epm{\end{pmatrix}}

\newcommand {\apgt} {\ {\raise-.5ex\hbox{$\buildrel>\over\sim$}}\ }
\newcommand {\aplt} {\ {\raise-.5ex\hbox{$\buildrel<\over\sim$}}\ }

\newcommand{\si}{\sigma}

\newcommand{\barray}{\begin{eqnarray}}
\newcommand{\earray}{\end{eqnarray}}
\newcommand{\nn}{\nonumber}
\newcommand{\disp}[1]{Eq.~(\ref{#1})}

\newcommand{\refdisp}[1]{Ref.~(\onlinecite{#1})}
\newcommand{\figdisp}[1]{Fig.~(\ref{#1})}

 \definecolor{indigo}{rgb}{0.5,0.1,1}

\begin{document}

\title{Band-edge quasiparticles from electron phonon coupling and resistivity saturation. }

\author{Edward Perepelitsky}
\affiliation{Physics Department, University of California,  Santa Cruz, CA 95064    }
\author{B. Sriram Shastry}
\affiliation{Physics Department, University of California,  Santa Cruz, CA 95064  }

\begin{abstract}

We address the problem of resistivity saturation observed in materials such as the A-15 compounds. To do so,
we calculate the resistivity for the Hubbard-Holstein model in infinite spatial dimensions to second order in on-site repulsion $U\leq D$ and to first order in (dimensionless) electron-phonon coupling strength $\lambda\leq0.5$, where $D$ is the half-bandwidth. We identify a unique mechanism to obtain two parallel quantum conducting channels: low-energy and band-edge high-energy quasi-particles. We identify the source of the hitherto unremarked high-energy quasi-particles as a positive slope in the frequency-dependence of the real part of the electron self-energy. In the presence of phonons, the self-energy grows linearly with the temperature at high-$T$, causing the resistivity to saturate. As $U$ is increased, the saturation temperature is pushed to higher values, offering a mechanism by which electron-correlations destroy saturation.
\end{abstract}

\pacs{71.10.Fd,71.27.+a,72.15.-v,72.15.Lh}

\maketitle

\section{Introduction}

The resistivity has been observed to saturate at high-temperatures in certain materials, such as the A-15 compounds, while growing without bound in others, such as the cuprates \cite{Gunnarson-Calandra, Fisk-Webb, Mooij, Kakeshitaetal, ando,Martin-Gurvitch, Takagi, Hebardetal}. Resistivity saturation has been seen as a signature of electron-phonon interactions \cite{Gurvitch1987} and weak electron-electron interactions. Many theoretical mechanisms have been proposed to address the problem of resistivity saturation \cite{Gunnarson-Calandra, Yu_Anderson, Gurvitch1981, Wermanetal, Millis, Pickett, Wiesmannetal, PBAllen, Laughlin}. In this paper, we offer a unique mechanism: the presence of two parallel quantum conducting channels consisting of the usual low-energy and the less obvious high-energy quasi-particles. These emergent objects derive from electron-phonon interactions. This is the main idea of our work, namely the role of the hitherto unnoticed high-energy (i.e. band-edge) quasiparticles, residing at or beyond the edge of the bare band. The demonstration of this idea requires only low order perturbation theory.

In particular, we evaluate the bare diagrams to leading order in the electron-phonon coupling. It would also have been possible to re-sum an infinite subset of diagrams by doing a self-consistent version of the same approximation. Moreover, for low energies, these are the only diagrams which contribute (Migdal's theorem) \cite{Migdal}. However, in our work, it is in fact the high-energy quasiparticles which play a key role, and therefore the use of Migdal's theorem is no longer justified. Therefore, all higher order diagrams enter into the series on equal footing. In the case of weak electron-phonon coupling, the approximation used here is rigorously justified, while for the case of intermediate or strong-coupling, we consider it to be the most unbiased. It has also been shown in recent work that self-consistent diagrammatic approximations can lead to wrong results in certain cases \cite{Koziketal}. Due to the nature of our approximation, we restrict $\lambda\leq0.5$.

In the presence of phonons, the high-energy quasi-particles lead to resistivity saturation. The mechanism we propose has a unique signature in the LDOS, which acquires peaks at or beyond the edge of the bare band. Therefore, it can be identified experimentally using ARPES/STM measurements. It also has a distinct signature in the optical conductivity (see SM-III), and can therefore be identified using the latter as well.

\begin{figure}[h]
\begin{center}
\includegraphics[width=.65 \columnwidth]{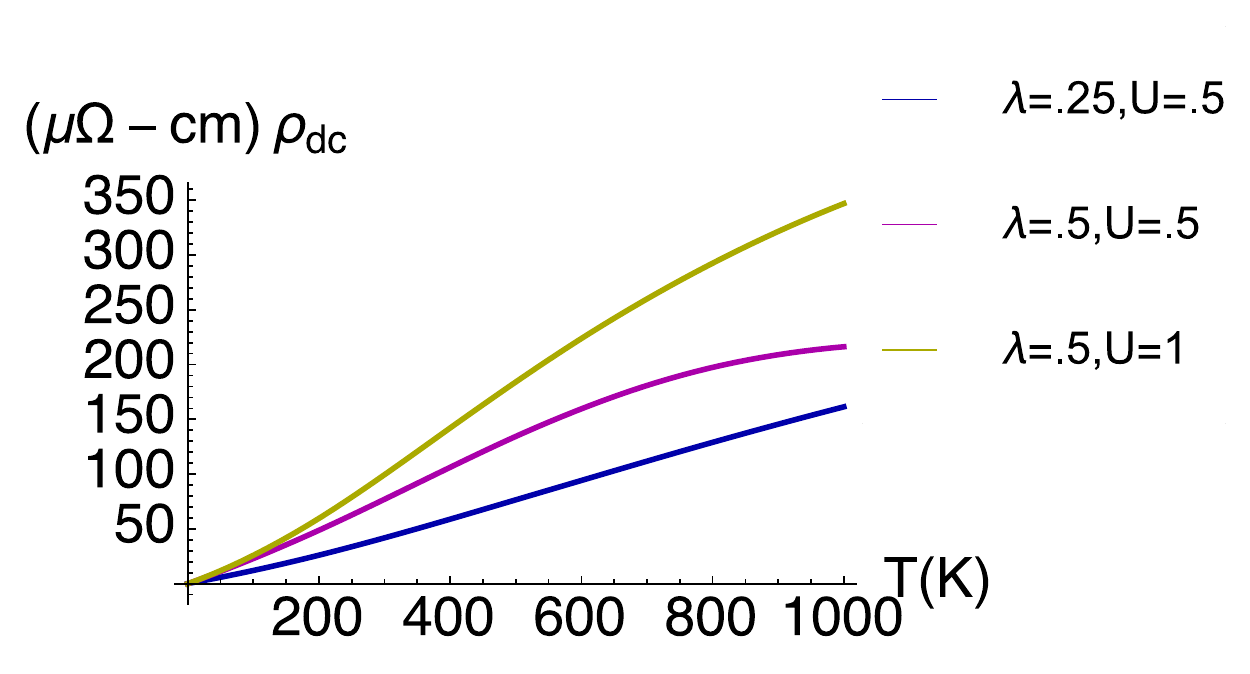}
\caption{ The dc resistivity vs. temperature for the Hubbard-Holstein model for three sets of parameters: ($\frac{U}{D}=.5$;$\lambda=.25$), ($\frac{U}{D}=.5$;$\lambda=.5$), and ($\frac{U}{D}=1$;$\lambda=.5$), with $D$ estimated as $2000$K. The prolonged region of negative curvature found in the middle set is observed in the A-15 compounds $Nb_3Sn$ and $Nb_3Sb$ \cite{Fisk-Webb}.}
\label{resexp}
\end{center}
\end{figure}

We study the Hubbard-Holstein model on the Bethe lattice in the limit of infinite spatial dimensions. The electrons interact through on-site repulsion $U$, and couple to an Einstein phonon mode with dimensionless electron-phonon coupling strength $\lambda$. We perform perturbation theory to second order in $U$ and to first order in $\lambda$. We compute the dc resistivity over a large range of temperature. We find that it displays resistivity saturation. In \figdisp{resexp}, we plot the resistivity, $\rho$, measured in units of $\mu\Omega$-$cm$, as a function of the temperature, $T$, for $T\leq 1000 K$. We computed this resistivity for three sets of parameters: ($\frac{U}{D}=.5$;$\lambda=.25$), ($\frac{U}{D}=.5$;$\lambda=.5$), and ($\frac{U}{D}=1$;$\lambda=.5$)  where $D$ is the half-bandwidth. These values seem compatible with  the  perturbative scheme employed.
 For the middle set of parameters, the calculated resistivity happens to be in good quantitative agreement with the resistivity observed in the A-15 compounds $Nb_3Sn$ and $Nb_3Sb$ \cite{Fisk-Webb}. In particular, note the negative curvature of the resistivity vs. temperature curve for $T\gtrsim 500 K$. For either weaker electron-phonon coupling or stronger electron-electron interactions, the saturation temperature increases beyond the scale that is probed in experiments.

\FloatBarrier

While previous studies of transport have focused on the conduction of low-energy quasiparticles, we identify a parallel quantum conduction channel, consisting of high-energy quasiparticles defined by   peaks in the spectral function, at, or beyond, the edge of the bare band that are sharp enough to be identifiable under certain conditions. 
In contrast to the low-energy quasi-particles, which are scattered more strongly at higher temperatures, the high-energy quasi-particles are pushed to higher energies with increasing temperature, and are therefore scattered more weakly. Denoting the resistivity of the low- and high-energy quasiparticles by $\rho_{ideal}$ and $\rho_{sat}$, respectively, the overall resistivity is given by the parallel resistor formula 
\beq
\frac{1}{\rho} = \frac{1}{\rho_{ideal}}+\frac{1}{\rho_{sat}},
\label{parallelresistors}
\eeq
 where $\rho_{ideal}$ ($\rho_{sat}$) is defined at all temperatures as the low (high)-frequency contribution to the integral in \disp{conductivitydefn-1} (detailed in supplementary materials Eqs.~(SM-1,SM-2)).

\begin{figure}[h]
\begin{center}
\includegraphics[width=.65 \columnwidth]{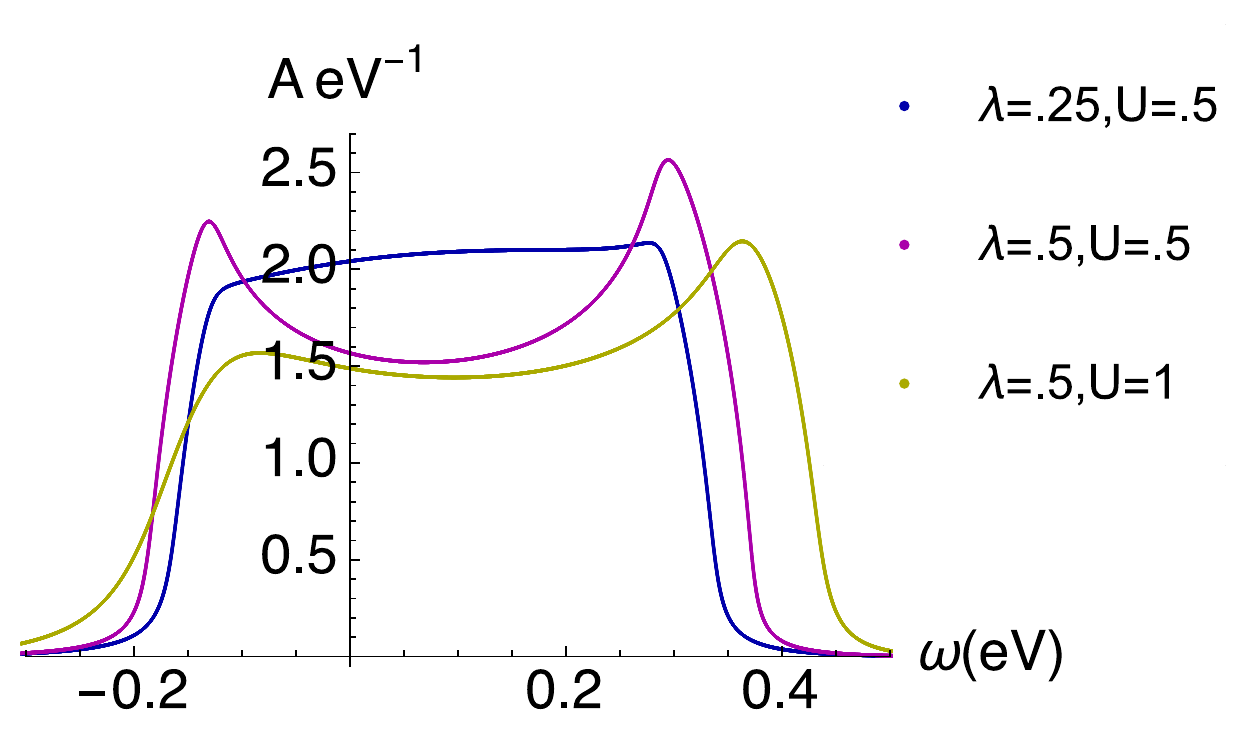}
\caption{ The LDOS at $T=1000K$ for the same sets of parameters as in \figdisp{resexp}. The peaks located beyond the edges of the bare band are signatures of the high-energy quasiparticles, and are a prediction of our theory which can be tested using STM. Figs. 4,8 display the T variation of $A$.}
\label{LDOSexp}
\end{center}
\end{figure}

As the temperature increases, the high-energy channel short circuits the low energy channel, and the resistivity saturates. The high-energy quasi-particles are visible in the local density of states (LDOS), which develops peaks at high-energies as the temperature is increased, while the central peak, associated with the low-energy quasiparticles, simultaneously shrinks. This is a prediction of our theory which can be tested using Scanning Tunneling Microscopy (STM). In \figdisp{LDOSexp}, we plot the LDOS, $A(\omega)$, at $T=1000 K$ for the same parameters as used in \figdisp{resexp}. The high-energy peaks are diminished by either decreasing $\lambda$ or by increasing $U$.

\FloatBarrier

\section{The model and calculation}

The Hamiltonian for our model containing disorder, interactions and a local Einstein mode phonon \cite{GirvinJonson} is the following:

\barray
H &=& \sum_k(\varepsilon_k-\mu )a^\dagger_ka_k + \omega_0 \sum_q b_q^\dagger b_q + U\sum_i n_{i\uparrow}n_{i\downarrow} \nn\\
&& + \frac{g}{\sqrt{N_s}}\sum_{k,q} a^\dagger_{k+q}a_k(b_q + b_{-q}^\dagger) + \sum_j \varepsilon_j n_{j \si},  
\earray
where $a_k$ is the electron destruction operator in momentum state $k$, $\varepsilon_k$ is the dispersion of the lattice, $b_q$ is the phonon destruction operator in momentum state $q$, $\omega_0$ is the energy of all phonon modes, $U$ is the on-site Hubbard repulsion, $N_s$ is the number of sites in the lattice, and $g$ is the electron-phonon coupling energy.
The $\varepsilon_j$ are quenched random site energies, which are treated within the Born approximation \cite{Abrikosov-Gorkov}, whereby the impurity averaged non-interacting electron Greens function is broadened $G_0^{-1} \to G_0^{-1} + i \,  \eta$
and $\eta= n_i \pi {\cal D}(\varepsilon_F) \langle \varepsilon_j^2\rangle $.

The electrons hop on the infinite-dimensional Bethe lattice, which has the density of states for energy  $\varepsilon\in[-D,D]$
\beq
{\cal D}(\varepsilon) = \frac{2}{\pi D} \left[1-\left(\frac{\varepsilon}{D}\right)^2\right]^{1/2} \label{DOS}
\eeq
where $D$ is the half-bandwidth. For the remainder of the paper, all energies will be measured in units of $D\sim 2000K$ \cite{Mattheiss-Weber}.

Following \refdisp{GirvinJonson} (see supplementary materials for details), for $T\gtrsim \omega_0$, the electron-phonon self-energy, computed to $O(g^2)$, is expressed as:
 
\beq
\rho_{\Sigma_{el,ph}}(\omega-\mu_{el}) = \frac{\pi\lambda A_{el}(\omega-\mu_{el})}{2} \times T,
\label{rhoSigapprox}
\eeq
where $\lambda$, defined by $g^2\equiv\frac{\pi D \lambda \omega_0}{4}$, is a dimensionless measure of the electron-phonon coupling strength. For any dynamical object $Q(\omega)$, $\rho_Q(\omega)\equiv -\frac{1}{\pi} \Im m (Q(\omega))$, and the subscript ``el" refers to quantities computed in the absence of phonons (g=0), using second order perturbation theory in the Hubbard $U$. Finally, $A(\omega)$ is the LDOS, obtained by integrating $\rho_G(\epsilon,\omega)$ over $\epsilon$, the latter obtained from $\Sigma(\omega)$ using Dyson's equation.

The dc conductivity can be expressed in terms of the spectral function via the formula\cite{badmetals}:
\barray
\sigma &=& \frac{2}{\pi T}   \ \sigma_{IRM} \int d\omega f(\omega-\mu_{el})\bar{f}(\omega-\mu_{el}) I(\omega-\mu_{el}),\;\;\; \;\;\label{conductivitydefn-1}
\earray
where $f(\omega)\equiv \frac{1}{e^{\beta \omega} + 1}$, $\bar{f}(\omega)\equiv 1-f(\omega)$, and $\beta\equiv\frac{1}{T}$. The spectral intensity, $I(\omega)$, is defined as
\barray
I(\omega-\mu_{el})&=&\pi^2 \int d\epsilon  \ \phi(\epsilon) \rho^2_G(\epsilon,\omega-\mu_{el}),
\label{conductivitydefn}
\earray
where $ \sigma_{IRM} \equiv \frac{1}{\rho_{IRM}}$, $\rho_{IRM}$ is the Ioffe-Regel-Mott limit of the resistivity, and the transport function
is given explicitly as $ \phi(\epsilon)= \Theta(1-\epsilon^2) \times (1-\epsilon^2)^{\frac{3}{2}}$. We will measure the resistivity in units of $\rho_{IRM}\approx258\mu\Omega cm$\cite{cutoff}. In the limit that $\rho_\Sigma(\omega-\mu_{el}) \ll D$ \cite{badmetals,cutoff},
\beq
I(\omega-\mu_{el}) \approx \frac{1}{2} \frac{\phi\left[R(\omega-\mu_{el})\right]}{\rho_\Sigma(\omega-\mu_{el}) },
\label{Iapprox}
\eeq
where $\Sigma(\omega) = \Sigma_{el}(\omega) + \Sigma_{el,ph}(\omega)$,  $R(\omega-\mu_{el})\equiv \omega +\Delta\mu - \Re e \ \Sigma(\omega-\mu_{el})$, and  $\Delta\mu\equiv \mu-\mu_{el}$ tends to $0$ as $T\to\infty$. Both $\mu$ and $\mu_{el}$ are determined through the particle sum-rule for the Green's function. The approximation \disp{Iapprox} is excellent for the dominant frequency range at all temperatures.

In our treatment of \disp{conductivitydefn-1}, we eschew the popular Sommerfeld  approximation \cite{Bloch,Gruneisen} $f(\nu)\bar{f}(\nu) \to T \delta(\nu)$, since it misleadingly throws out the contribution from the high-energy peaks in $I(\nu)$. In \disp{Iapprox}, we note that the spectral intensity $I$ can be large at any frequency  where the imaginary self energy is small, and the real part of the inverse Green's function is also small.

\section{Results}

We choose the parameters as follows. We set the density to $n=0.7$. Since our calculation does not incorporate any Mott physics, the exact value of the density does not change the qualitative features of the results. We choose the phonon energy $\omega_0=0.015$. Since the temperature regime of interest is in the semi-classical regime ($T\gg\omega_0$), the value of $\omega_0$ has very little bearing on the results (see \disp{rhoSigapprox}). We choose the impurity scattering $\eta =0.0012$. $\eta$ is chosen to be small but finite to ensure that the resistivity does not abruptly drop to zero above a certain temperature. Once again, in the range of experimentally relevant temperatures, $\eta$ has little bearing on the results. Finally, we restrict $\lambda\leq0.5$ and $U\leq D$, so that low-order perturbation theory can be expected to give reliable results.

\subsection{$U=0$.}

\begin{figure}[h]
\begin{center}
\includegraphics[width=.7 \columnwidth]{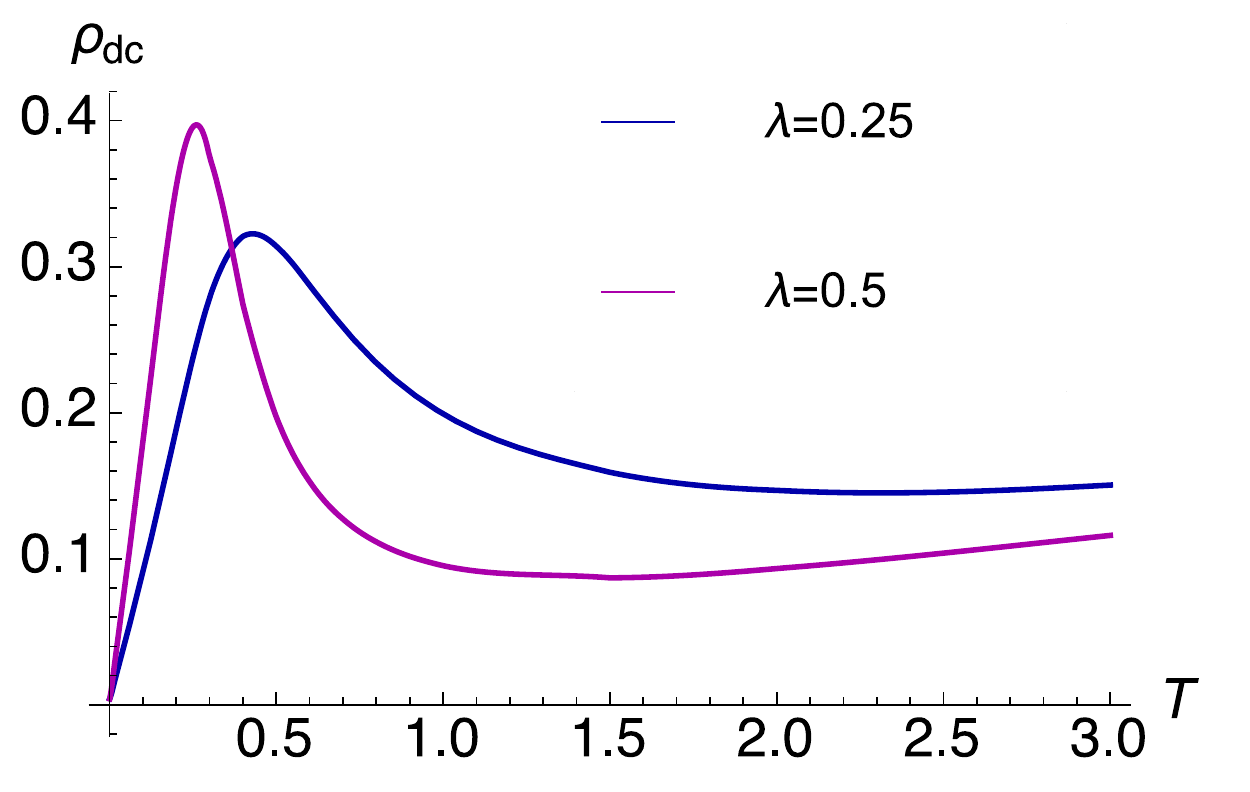}
\caption{$\rho_{dc}$ vs. $T$ for $\lambda =0.25, 0.5$ and $U=0$. As $\lambda$ increases, $T_{max}$ decreases, while the height of the peak increases. In the $T\to\infty$ limit, the resistivity curves collapse onto a straight line. }
\label{reslambda}
\end{center}
\end{figure}

In the case of $U=0$, the free electrons are scattered by phonons and impurities. In \figdisp{reslambda}, we plot the  resistivity for $\lambda =0.25$ and $\lambda = 0.5$. In both cases, the resistivity displays a maximum at $T\equiv T_{max}$, before finally increasing again at high-temperatures. As $\lambda$ increases, $T_{max}$ decreases, while the height of the peak increases. In the $T\to\infty$ limit, the resistivity curves collapse onto a straight line, whose slope is fixed by the impurity scattering of the electrons. This picture differs from the textbook discussions\cite{Ziman} of a monotonically increasing phononic resistivity. In the latter, the electrons are modeled as an electron gas with an infinite bandwidth, while here, the narrow electronic band is a key component, leading to high-energy quasiparticles at or beyond the edge of the bare band. The presence of the high-energy parallel conducting channel causes a rollover and hence a maximum in the curve. In fact, resistivity curves similar to those in \figdisp{reslambda} have also been found in theoretical studies of the Periodic Anderson Model and the Kondo Lattice model \cite{Schweitzer-Czycholl,Burdinetal,Pruschkeetal,Burdin-Zlatic,Grenzebachetal,Tahvildar-Zadehetal,Vidhyadhirajaetal}.

\begin{figure}[h]
\begin{center}
\includegraphics[width=.7 \columnwidth]{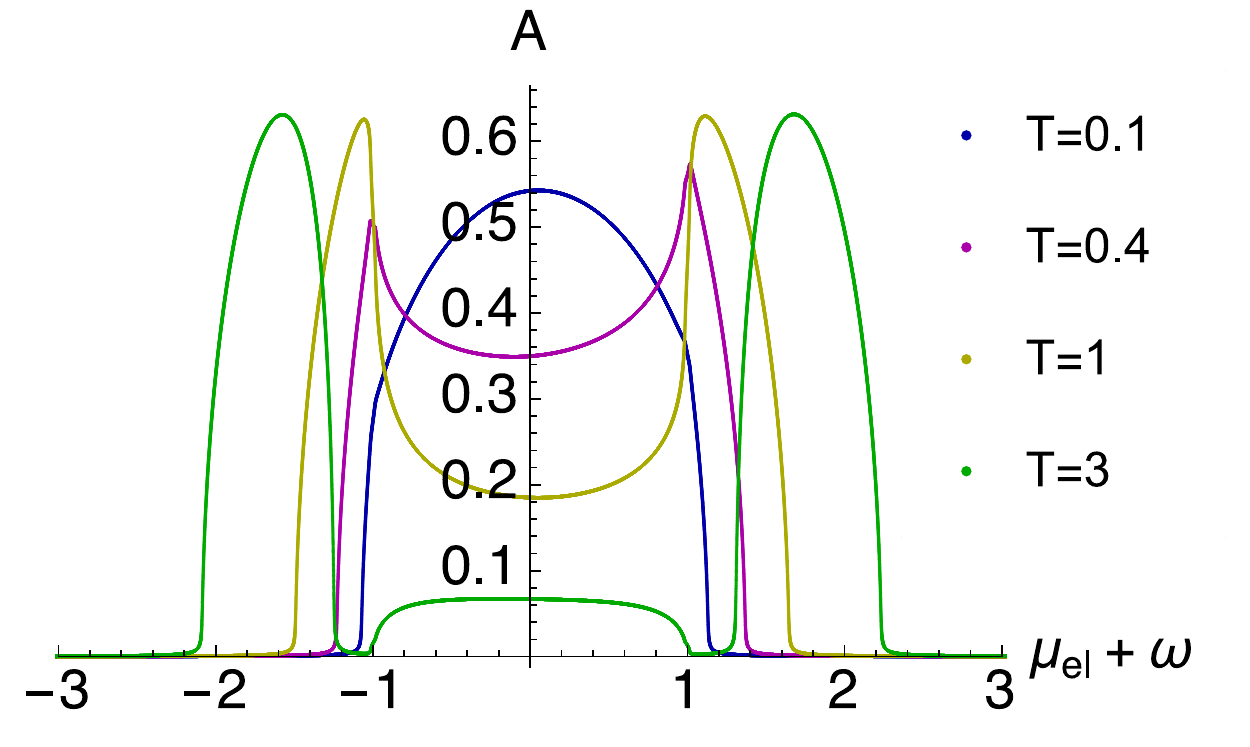}
\caption{The LDOS for $\lambda=0.5$ and $U=0$ at $T=0.1, \ 0.4, \ 1, \ 3$. For $T\lesssim T_{max}$, the central peak is a signature of the low-energy quasiparticles, while for $T\gtrsim T_{max}$, the two high-energy peaks are  signatures of the high-energy quasiparticles. The high-energy peaks get pushed to higher energies with increasing temperature. As the scattering rate of the high-energy quasiparticles decreases, so does the resistivity (\figdisp{reslambda}). }
\label{LDOSU0}
\end{center}
\end{figure}

In \figdisp{LDOSU0}, we plot the LDOS for $\lambda=0.5$ at $T=0.1, \ 0.4, \ 1, \ 3$. For $T\lesssim T_{max}$, the LDOS consists of a single central peak, and hence the conductivity is dominated by the low-energy channel ($\rho\approx \rho_{ideal}$). For $T\gtrsim T_{max}$, the LDOS consists of two high-energy peaks, and hence the conductivity is dominated by the high-energy channel ($\rho\approx \rho_{sat}$). As the temperature increases past $T_{max}$, the high-energy quasi-particles are pushed to increasingly higher energies and have correspondingly smaller scattering rates, causing the resistivity to decrease.

\begin{figure}[h]
\begin{center}
\includegraphics[width=.7 \columnwidth]{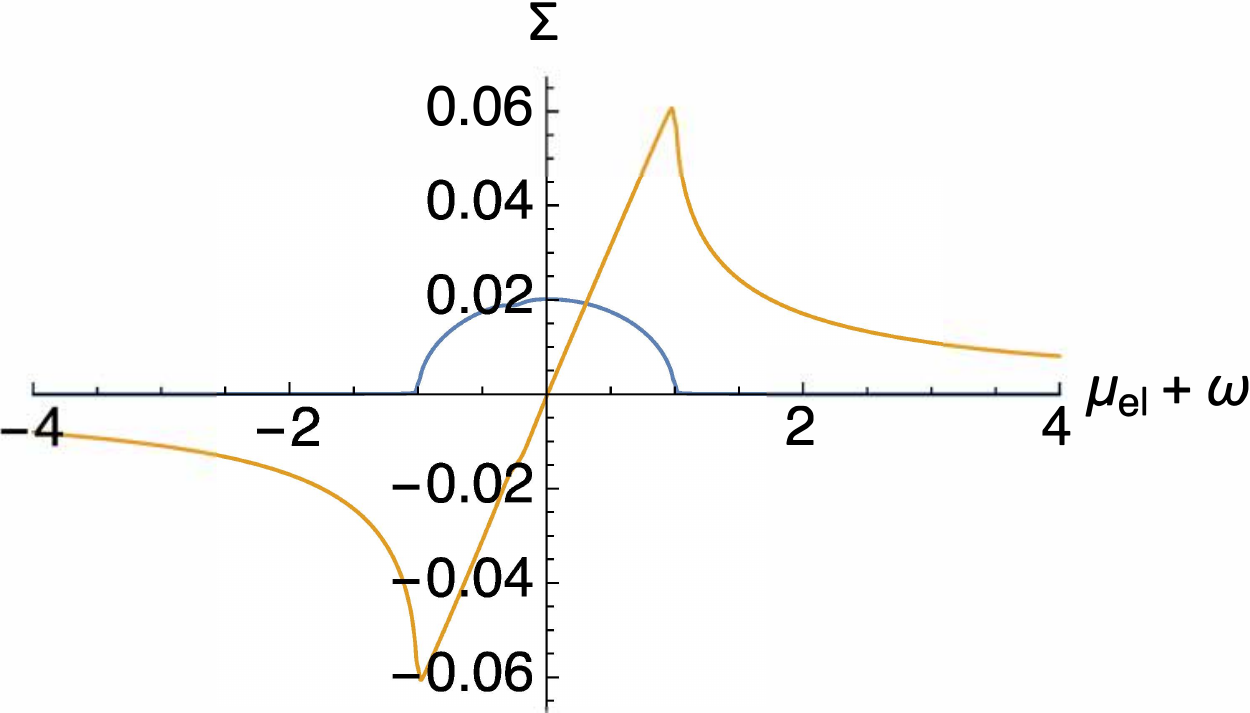}
\caption{ $\rho_\Sigma$ and $\Re e \Sigma $ for $\lambda=0.5$ and $U=0$ at $T=0.04$. This relatively low temperature is already in the semi-classical regime ($T\gtrsim\omega_0$), where Eqs. (\ref{rhoSigapprox2}) and (\ref{ReSigapprox}) are excellent approximations. The positive slope of $\Re e \Sigma $ is responsible for the high-energy quasiparticles (\figdisp{LDOSU0}).}
\label{selfenergy}
\end{center}
\end{figure}

The existence of high-energy quasiparticles requires $R(\omega-\mu_{el})$ to vanish at large values of the frequency. This in turn, requires that $\Re e \Sigma (\omega-\mu_{el})$ have positive slope of order unity. In \figdisp{selfenergy}, we plot $\rho_\Sigma$ and $\Re e \Sigma $ at $T=0.04$. Using \disp{rhoSigapprox} (for small $\eta$),

\beq
\rho_\Sigma(\omega-\mu_{el}) = \frac{\pi\lambda {\cal D}(\omega)}{2} \times T,
\label{rhoSigapprox2}
\eeq
\\
\beq
\Re e \ \Sigma(\omega-\mu_{el}) = \pi \lambda\times T \times \omega\left[1-\Theta(\omega^2-1)\sqrt{1-\frac{1}{\omega^2}}\right].
\label{ReSigapprox}
\eeq
The slope of $\Re e \ \Sigma(\omega-\mu_{el})$ increases linearly with $T$, pushing the high-energy quasiparticles to higher energies, causing their scattering rate to decrease. Meanwhile, the scattering rate of the low-energy quasiparticles grows linearly with $T$.

\FloatBarrier

\subsection{Finite-$U$.}

\begin{figure}[h]
\begin{center}
\includegraphics[width=.7 \columnwidth]{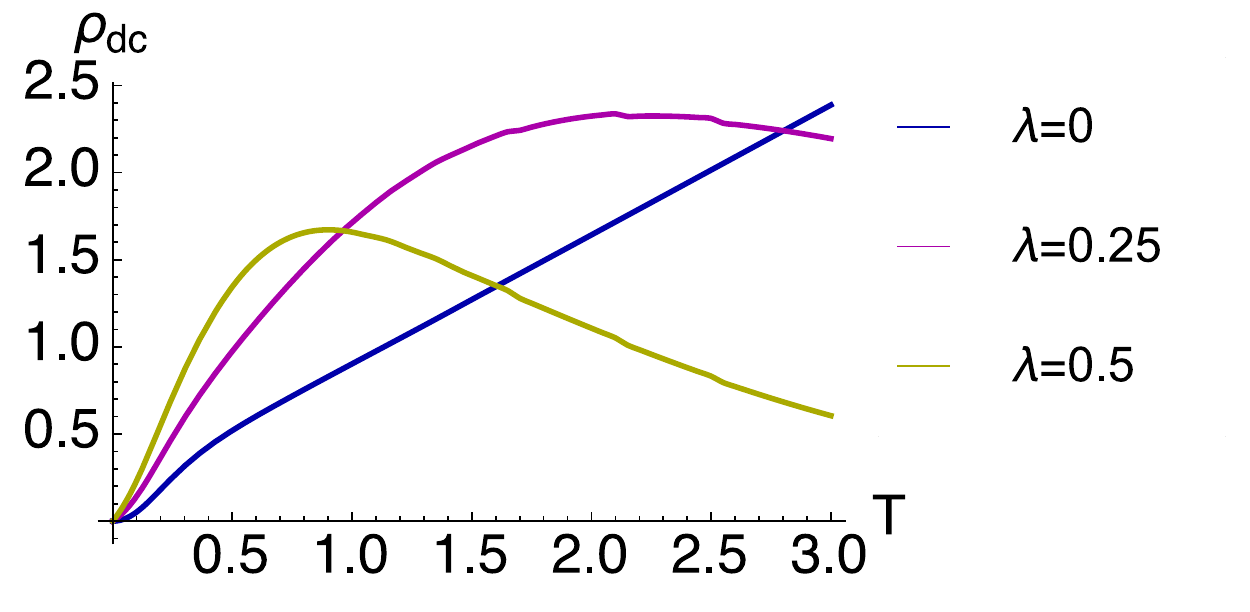}
\caption{ The resistivity vs. temperature curve for $\lambda =0,0.25, \ 0.5$ and $U=1$. For $\lambda\neq0$, the peak shifts to the left and becomes sharper with increasing $\lambda$.  Note that when $T$ is restricted to experimentally relevant values, i.e. $T\lesssim 0.5D$, only the left half of the peak appears, mimicking saturation (see \figdisp{resexp}). }
\label{resU1}
\end{center}
\end{figure}

\begin{figure}[h]
\begin{center}
\includegraphics[width=.7 \columnwidth]{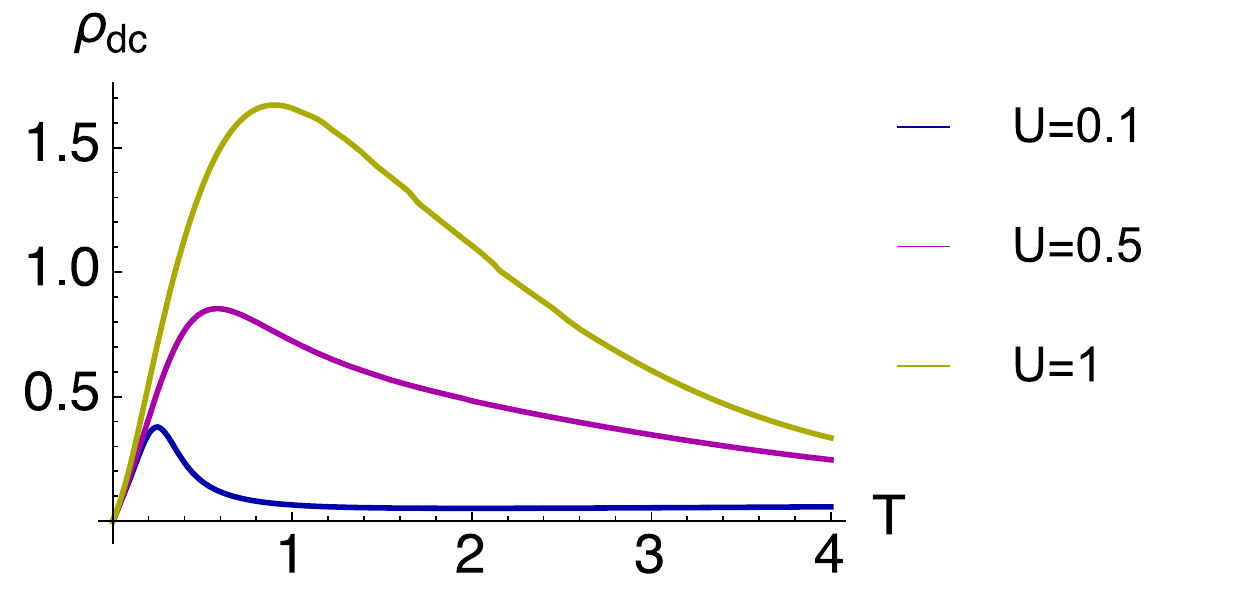}
\caption{The resistivity vs. temperature curve for $\lambda = 0.5$ and $U=0.1, \ 0.5, \ 1.0$. $T_{max}$ increases with increasing $U$.}
\label{resvaryU}
\end{center}
\end{figure}

In \figdisp{resU1}, we plot the resistivity vs. temperature curve for $\lambda =0, 0.25, \ 0.5$ and $U=1$. For $\lambda=0$, i.e. the Hubbard Model, the resistivity is monotonic, but has a kink at $T_{kink}\approx 0.4$. For $T_{FL}<T<T_{kink}$, the resistivity is quasi-linear with negative intercept, while $T>T_{kink}$, it is quasi-linear with positive intercept. Here, $T_{FL}$ is the Fermi-liquid scale, below which the resistivity is quadratic with temperature. These features are also observed in DMFT studies of the Hubbard model in both the cases of small and large $U$ \cite{GeorgesetalRMP,badmetals,wenxinetal, high-T}. For finite $\lambda$, the resistivity once again displays a maximum at $T_{max}$. $T_{max}$ decreases and the peak becomes sharper with increasing $\lambda$.  In \figdisp{resvaryU}, we plot the resistivity vs. temperature curve for $\lambda = 0.5$ and $U=0.1, \ 0.5, \ 1.0$. $T_{max}$ increases with increasing $U$. 

\begin{figure}[h]
\begin{center}
\includegraphics[width=.7 \columnwidth]{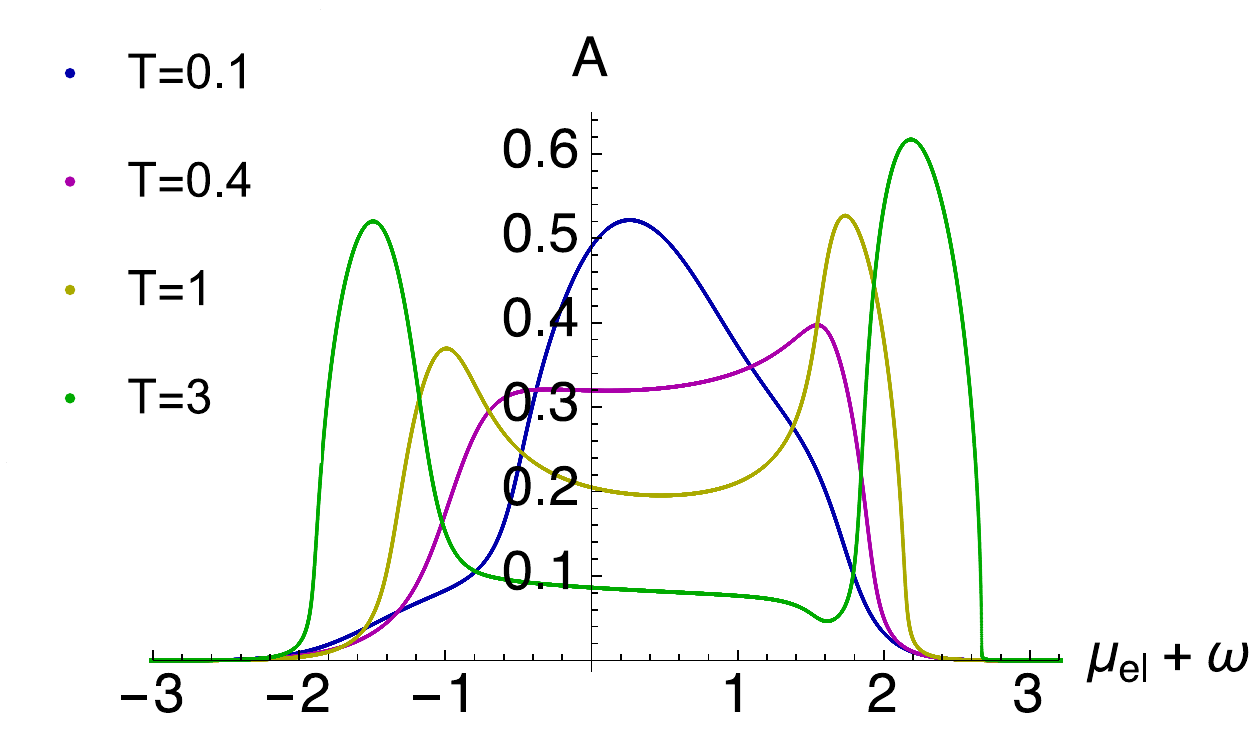}
\caption{The LDOS for $\lambda=0.5$ and $U=1$ at $T=0.1, \ 0.4, \ 1, 3$. For $T\lesssim T_{max}$, the central peak is a signature of the low-energy quasiparticles, while for $T\gtrsim T_{max}$, the high-energy peaks are a signature of the high-energy quasiparticles. }
\label{LDOSU1lamp5}
\end{center}
\end{figure}

In \figdisp{LDOSU1lamp5}, we plot the LDOS for $\lambda=0.5$ and $U=1$ at $T=0.1, \ 0.4, \ 1, 3$. For $T\lesssim T_{max}$, the central peak is a signature of the low-energy quasiparticles, while for $T\gtrsim T_{max}$, the high-energy peaks are a signature of the high-energy quasiparticles. Comparing with \figdisp{LDOSU0}, the high-energy peak values are smaller. This is a consequence of the broadening of the imaginary part of the self-energy beyond the edges of the bare-band (see \figdisp{selfenergy} and SM-Fig.(1)) as $U$ is increased. Consequently, the conductivity in the high-energy channel decreases, and $T_{max}$ shifts to the right with increasing $U$ (see \figdisp{resvaryU}).  

\section {Conclusion} 

We have computed the resistivity vs. temperature curve in the Hubbard-Holstein model on the infinite dimensional Bethe lattice, with weak  to intermediate electronic repulsion $U\leq D$, and  electron-phonon coupling strength $\lambda\leq 0.5$. For $\lambda>0$, it has a broad maximum, consistent with materials that display resistivity saturation. For $\lambda=0$, it has a kink rather than a maximum. 

We have identified two parallel quantum conducting channels, consisting of low- and high-energy quasiparticles. The former dominates at low temperatures, causing the resistivity to increase, while the latter dominates at high temperatures, causing the resistivity to saturate. The temperature scale of the saturation increases with both increasing $U$ and decreasing $\lambda$.

Finally, we have traced the origin of the high-energy quasiparticles to the frequency-dependence of $ \Re e \Sigma$, which must have a region of positive slope of order unity. This shape is inherited from the Hilbert transform of the electronic LDOS (with $\lambda$=0, see \disp{rhoSigapprox}). 

It is possible that resistivity saturation can be achieved by more than one mechanism. The mechanism which we propose (i.e. high energy quasiparticles) has a distinct signature in the LDOS, which can be observed using ARPES/STM, as well as in the optical conductivity.

\section{Acknowledgements}

We would like to thank Antoine Georges for many stimulating and enlightening discussions. We would like to thank the referees for helpful comments. This work was supported by the US Department of Energy (DOE), Office of Science, Basic Energy Sciences
(BES), under Award No. DE-FG02-06ER46319.



%

\end{document}


\title{ Supplementary Material:\\
Band-edge quasiparticles from electron phonon coupling and resistivity saturation.}
\author{Edward Perepelitsky}
\affiliation{Physics Department, University of California,  Santa Cruz, CA 95064    }
\author{B. Sriram Shastry}
\affiliation{Physics Department, University of California,  Santa Cruz, CA 95064  }

\maketitle
\makeatletter
\renewcommand\@biblabel[1]{[SI-#1]}
\makeatother

\section{Self-energy in Finite-$U$ case.}

\begin{figure}[h]
\begin{center}
\includegraphics[width=.7 \columnwidth]{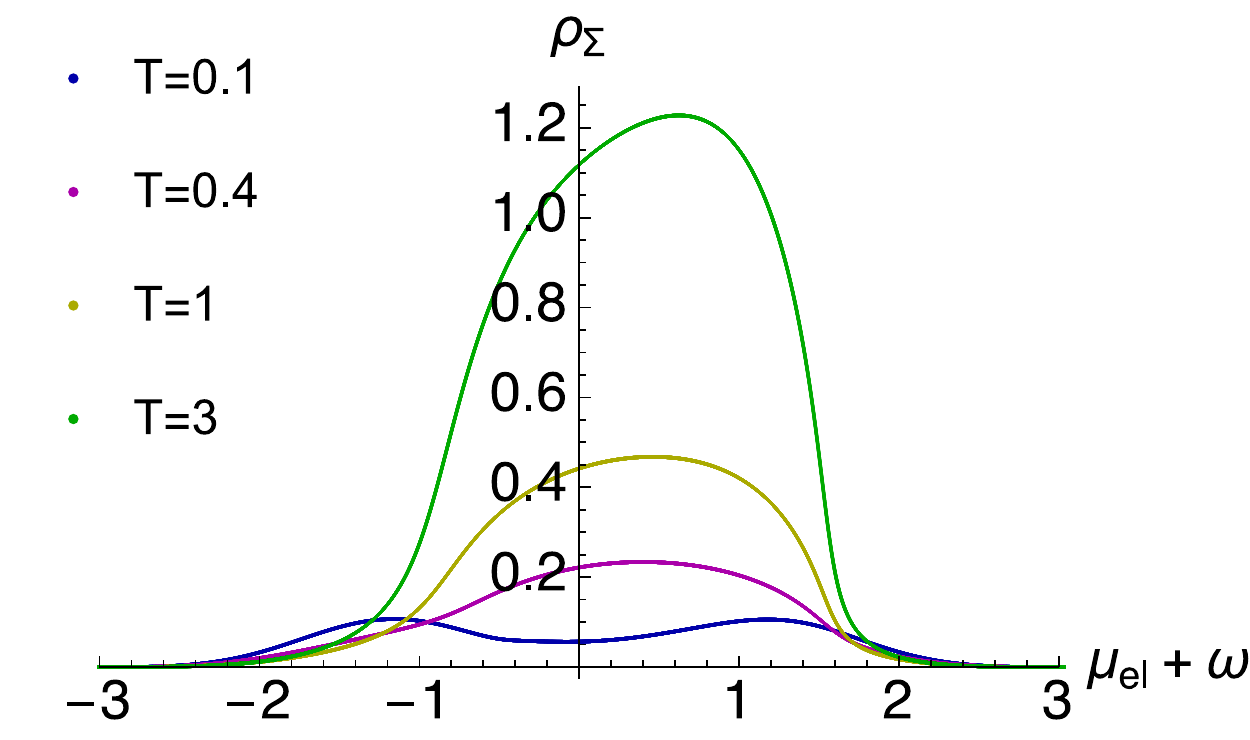}
\caption{$\rho_{\Sigma}$ for $U=1$ and $\lambda=0.5$ at $T=0.1, \ 0.4, \ 1, \ 3$.  Compared with the $U=0$ case (see Fig.(5) of the main paper), the tails spill over beyond the edges of the bare band. }
\label{rhoSigU1lamp5}
\end{center}
\end{figure}

\begin{figure}[h]
\begin{center}
\includegraphics[width=.7 \columnwidth]{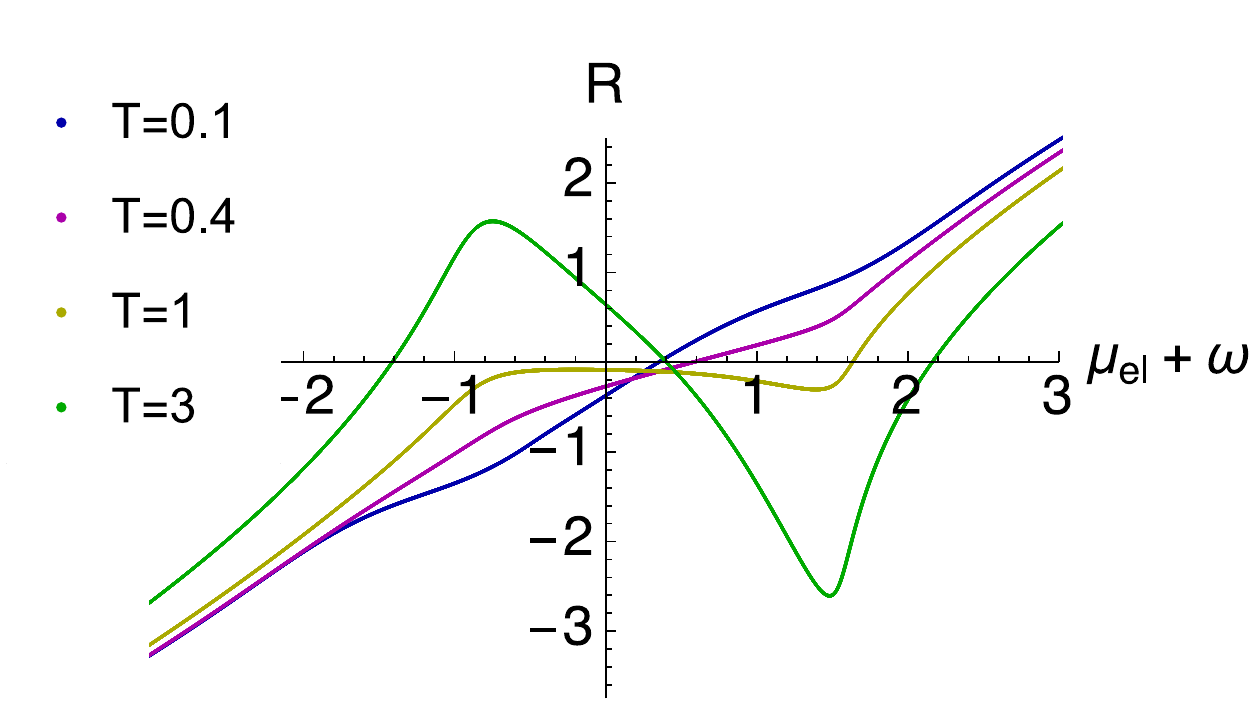}
\caption{ $R$ for $U=1$ and $\lambda=0.5$ at $T=0.1, \ 0.4, \ 1, \ 3$. As the temperature is raised, $R(\omega - \mu_{el})$ bends towards the frequency axis at high-energies, leading to the high-energy quasiparticles.}
\label{RU1lamp5}
\end{center}
\end{figure}

We consider once again $U=1$. In \figdisp{rhoSigU1lamp5}, we plot $\rho_{\Sigma}(\omega-\mu_{el})$ for $\lambda=.5$ at $T=0.1, \ 0.4, \ 1, \ 3$.  Just as in the case of $U=0$, it consists of a peak centered near the origin, whose height grows in magnitude linearly with $T$ (see Fig.(5) of the main paper). However, in the finite-$U$ case, the tails spill over beyond the edges of the bare band, causing the resistivity to increase in the high-energy channel, and $T_{max}$ to increase with increasing $U$. The corresponding $R(\omega-\mu_{el})$ is plotted in \figdisp{RU1lamp5}.  The maximum in  $\rho_{\Sigma}(\omega-\mu_{el})$  causes  $\Re   e  \ \Sigma (\omega-\mu_{el})$ to have a broad region of positive slope. The resulting $R(\omega-\mu_{el})$ bends towards the frequency-axis at high-energies, leading to the high-energy quasiparticles. 
 
\section{Conductivity in the low and high frequency channels.}

Rewriting Eq.(1) in terms of conductivities, we have

\beq
\si = \si_{ideal}+\si_{sat},
\eeq
where $\si_{ideal}$ is the conductivity of the low-energy quasiparticles and $\si_{sat}$ is the conductivity of the high-energy quasiparticles. We can formally define each one as the low- and high- frequency contributions to the integral in Eq.(5):

\barray
\frac{\sigma_{ideal}}{\sigma_{IRM} } &=& \frac{2}{\pi T}   \ \int_{|\omega|<\omega_c} d\omega f(\omega-\mu_{el})\bar{f}(\omega-\mu_{el}) I(\omega-\mu_{el}),\nn\\
\frac{\sigma_{sat}}{\sigma_{IRM} } &=& \frac{2}{\pi T}   \ \int_{|\omega|>\omega_c} d\omega f(\omega-\mu_{el})\bar{f}(\omega-\mu_{el}) I(\omega-\mu_{el}).\nn\\
\earray

\begin{figure}[h]
\begin{center}
\includegraphics[width=.7 \columnwidth]{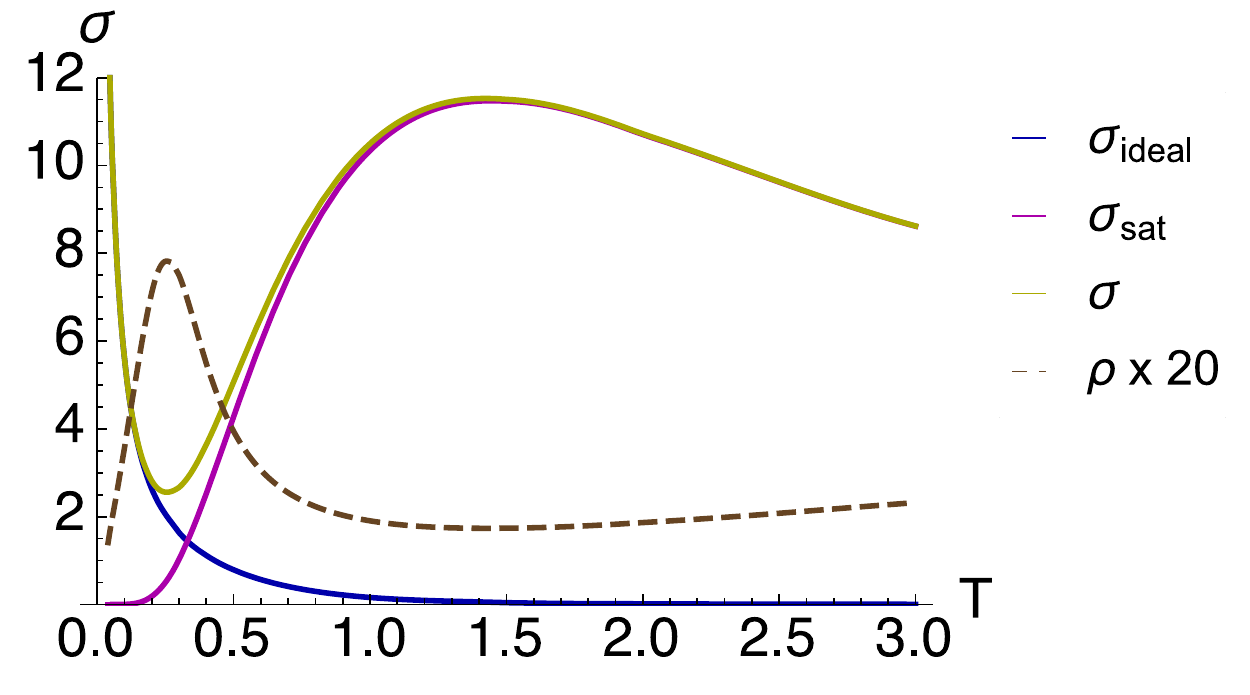}
\caption{ $\sigma_{ideal}$, $\sigma_{sat}$, and $\sigma$ as a function of $T$ for the case of $U=0$, $\lambda=0.5$. At low temperatures, $\sigma$ is dominated by $\sigma_{ideal}$, while at high temperatures, it is dominated by $\sigma_{sat}$. The crossover occurs at $T\approx T_{max}$, the location of the maximum in the resulting resistivity vs. temperature curve (plotted as the dashed brown curve (scaled by a factor of 20)).}
\label{siggtrsmU0}
\end{center}
\end{figure}

\begin{figure}[h]
\begin{center}
\includegraphics[width=.7 \columnwidth]{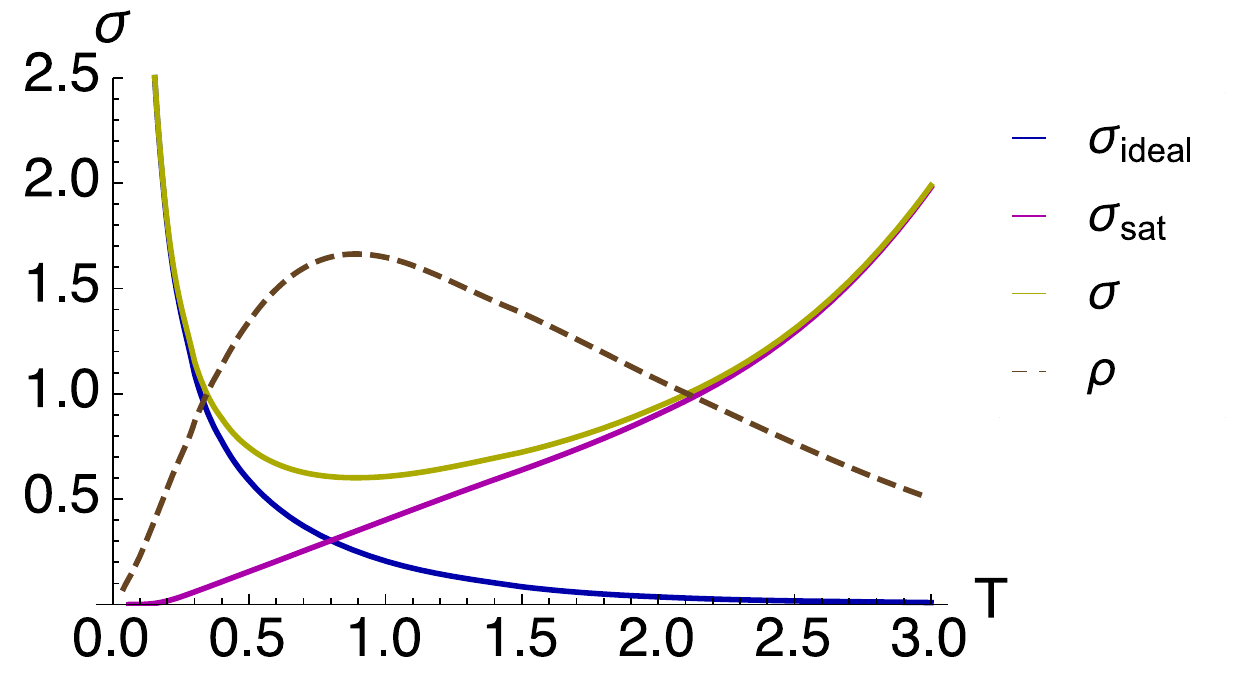}
\caption{ $\sigma_{ideal}$, $\sigma_{sat}$, and $\sigma$ as a function of $T$ for the case of $U=1$, $\lambda=0.5$. At low temperatures, $\sigma$ is dominated by $\sigma_{ideal}$, while at high temperatures, it is dominated by $\sigma_{sat}$. The crossover occurs at $T\approx T_{max}$, the location of the maximum in the resulting resistivity vs. temperature curve (plotted as the dashed brown curve). }
\label{siggtrsmU1lamp5}
\end{center}
\end{figure}

We choose $\omega_c = 0.95$ to be slightly smaller than the edge of the bare band to capture the entire high-energy quasiparticle in $\sigma_{sat}$. In Figs. (\ref{siggtrsmU0}) and (\ref{siggtrsmU1lamp5}), we plot $\sigma_{ideal}$, $\sigma_{sat}$, and $\sigma$ as a function of $T$ for the cases of $U=0$; $\lambda=0.55$, and $U=1$; $\lambda=0.5$, respectively. At low temperatures, $\sigma$ is dominated by the low-energy quasiparticles, while at high temperatures, it is dominated by the high energy quasiparticles. In both cases, the resulting resistivity vs. temperature curve has a maximum at approximately the crossover temperature.

\section{Optical Conductivity}

\begin{figure}[h]
\begin{center}
\includegraphics[width=.7 \columnwidth]{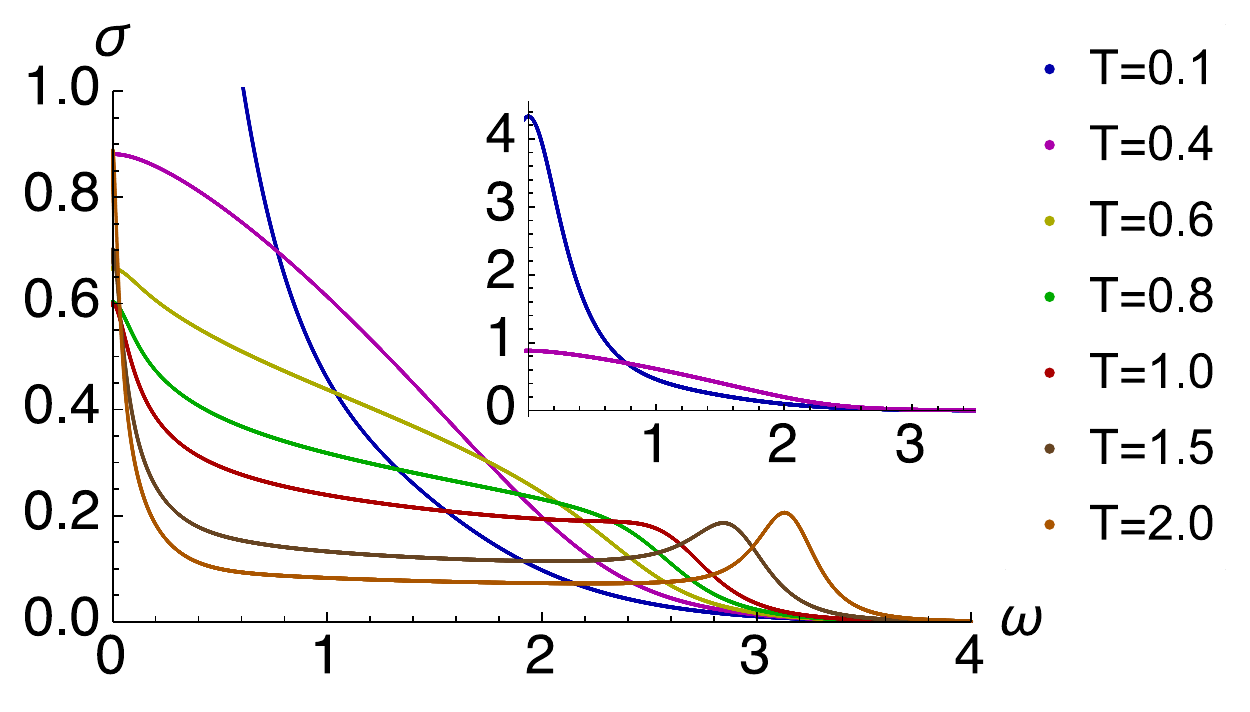}
\caption{The optical conductivity for $\lambda=0.5$ and $U=1$, at $T=0.1,0.4,0.6,0.8,1.0,1.5,2.0$. For $T\gtrsim T_{max}$, it develops a high-frequency feature, which both grows and moves to the right with increasing temperature. This feature corresponds to optical transitions between the high-energy quasiparticle peaks in the spectral function. Therefore, the high-energy quasiparticles have a distinct and observable signature in the optical conductivity.  }
\label{optcond}
\end{center}
\end{figure}

The high-energy quasi-particles have a distinct signature in the temperature-dependence of the optical conductivity (\disp{optconddefn}). In \figdisp{optcond}, we plot the optical conductivity for $\lambda=0.5$ and $U=1$, at $T=0.1,0.4,0.6,0.8,1.0,1.5,2.0$. The Drude peak, which is present at all temperatures, corresponds to zero-energy optical transitions between the central peak (in the spectral function) and itself at low temperatures, and the high-energy peaks and themselves at high-temperatures (see Fig.8 of MS). For $T\gtrsim T_{max}$ (see Fig.6 of MS), the optical conductivity develops a feature at high-frequencies, which both grows and moves to the right with increasing temperature. This corresponds to optical transitions between the high-energy peaks, which are growing and moving further apart with increasing temperature (see Fig.8 of MS).

\section {Dependence of $T_{max}$ on $\lambda$ and $U$.}

\begin{figure}[h]
\begin{center}
\includegraphics[width=.7 \columnwidth]{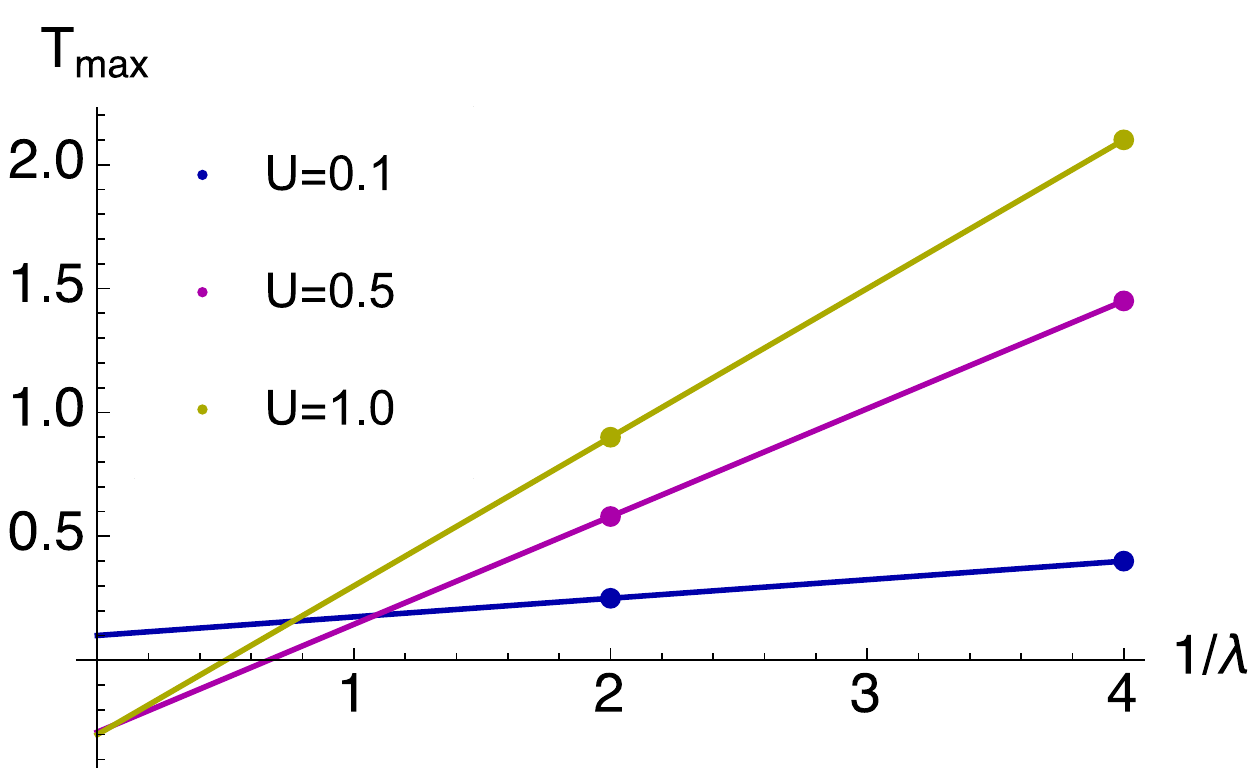}
\caption{ $T_{max}$ vs. $\frac{1}{\lambda}$ for $U = 0.1, \ 0.5, \ 1.0$, where $T_{max}$ is the peak location in the resistivity vs. temperature curve. It has the form $T_{max} = \frac{T_{max,s}}{\lambda} + T_{max,0}$, where $T_{max,0}$ and $T_{max,s}$ are undetermined fit parameters. }
\label{Tmax}
\end{center}
\end{figure}

\begin{figure}[h]
\begin{center}
\includegraphics[width=.7 \columnwidth]{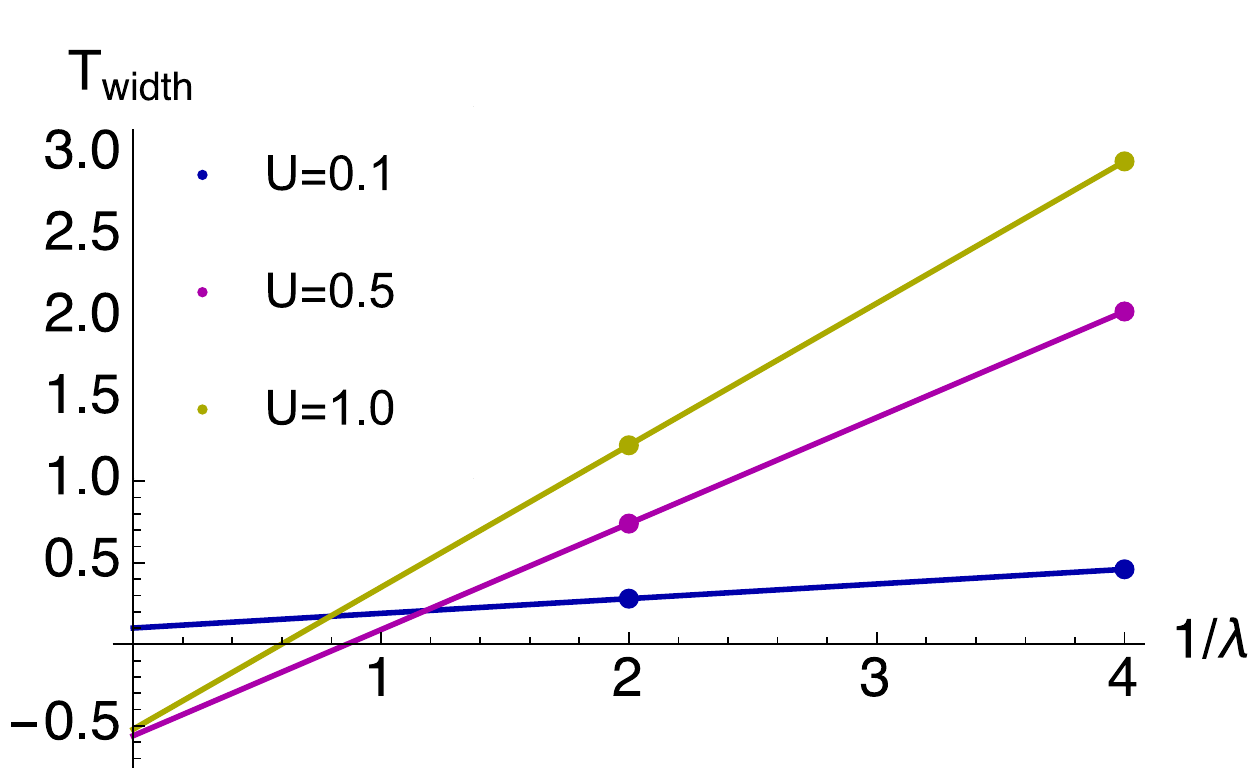}
\caption{ $T_{width}$ vs. $\frac{1}{\lambda}$ for $U = 0.1, \ 0.5, \ 1.0$, where $T_{width}$ is the peak width at half-maximum in the resistivity vs. temperature curve. It has the form $T_{width} = \frac{T_{width,s}}{\lambda} + T_{width,0}$, where $T_{width,0}$ and $T_{width,s}$ are undetermined fit parameters. }
\label{Twidth}
\end{center}
\end{figure}

In this section, we explore the dependence of $T_{max}$ on $\lambda$ and $U$. Note that in the equation for the self-energy (Eq.(4) of the main paper), $\lambda$ and $T$ appear together. In addition, the Fermi functions in the definition of the conductivity (Eq.(5) of the main paper) have explicit temperature dependence (the pre-factor of $\frac{1}{T}$ does not affect the value of $T_{max}$). Therefore in the $T_{max} \to \infty$ ($\lambda\to0$) limit, we expect $T_{max} \propto \frac{1}{\lambda}$. In \figdisp{Tmax}, plotting $T_{max}$ vs. $\frac{1}{\lambda}$ for $U = 0.1, \ 0.5, \ 1.0$ reveals that $T_{max} = \frac{T_{max,s}}{\lambda} + T_{max,0}$, where $T_{max,0}$ and $T_{max,s}$ are undetermined fit parameters. In the $\lambda\to0$ limit, this reproduces the expected result. Additionally, in \figdisp{Twidth}, we plot $T_{width}$ vs. $\frac{1}{\lambda}$ for $U = 0.1, \ 0.5, \ 1.0$, where $T_{width}$ is the width at half-maximum of the peak in the resistivity vs. temperature curve. Once again, we find that $T_{width} = \frac{T_{width,s}}{\lambda} + T_{width,0}$, where $T_{width,0}$ and $T_{width,s}$ are undetermined fit parameters. We summarize the values of the fit parameters for various values of $U$ in Table I.

\begin{table}
\begin{center}
\caption{Fit Parameters}
 \begin{tabular}{||c c c c c||} 
 \hline
 $U$ & $T_{max,0}$ &$T_{max,s}$ & $T_{width,0}$ &$T_{width,s}$\\ [0.5ex] 
 \hline\hline
 0.1 & 0.1 & 0.075 & 0.1 & 0.09 \\ 
 \hline
 0.5 & -0.29 &0.435 & -0.56 & 0.65 \\
 \hline
 1 & -0.3 & 0.6 & -0.52 & 0.87 \\ [1ex] 
 \hline
\end{tabular}
\end{center}
\end{table}

\section{Details of perturbative calculation.}

To compute $G(\varepsilon,\omega,\mu)$, the single-electron propagator, to $O(g^2)$, we use the following procedure. We first compute $\Sigma_{el}(\omega)$, the electron-electron scattering in the absence of phonons. The resulting single-electron propagator, $G_{el}(\varepsilon,\omega,\mu_{el})$, then scatters off of the phonons, producing $\Sigma_{el,ph}(\omega)$. $\mu_{el}$ is chosen to ensure the correct number sum rule for the electrons in the absence of the phonons. The total scattering of the electron, $\Sigma(\omega)$, is given by the sum: $\Sigma(\omega) = \Sigma_{el}(\omega) + \Sigma_{el,ph}(\omega)$. $G(\varepsilon,\omega,\mu)$ is given in terms of $\Sigma(\omega)$ using the usual Dyson's equation. $\mu$ is then chosen to ensure the correct number sum rule for the electrons in the presence of the phonons. Since $\mu - \mu_{el}$ is $O(g^2)$, it is consistent to replace $\mu$ by $\mu_{el}$ in $G_{el}$.

Using second order perturbation theory in $U$, we find that
\barray
&&\rho_{\Sigma_{el}}(\omega-\mu_{el}) = U^2 \int d\nu_1d\nu_2 {\cal D}(\nu_1){\cal D}(\nu_2){\cal D}(\nu_1+\nu_2-\omega)\times\nn\\
&&\left[f(\nu_1-\mu_{el})f(\nu_2-\mu_{el})\bar{f}(\nu_1+\nu_2-\omega-\mu_{el})+f\leftrightarrow\bar{f} \ \right] + \eta,\nn\\
&&\Re e \Sigma_{el}(\omega) \equiv U \ \frac{n}{2} + \int d\nu  \ \frac{\rho_{\Sigma_{el}}(\nu)}{\omega-\nu},\nn\\
\earray
where for any dynamical object $Q(\omega)$, $\rho_Q(\omega)\equiv -\frac{1}{\pi} \Im m (Q(\omega))$, $f(\omega)\equiv \frac{1}{e^{\beta\omega}+1}$, $\bar{f}(\omega)\equiv 1 -f(\omega)$ , and $\beta\equiv \frac{1}{T}$. We find it convenient to shift the frequency by the chemical potential, since the latter grows linearly with $T$ at high-$T$. $\rho_{G_{el}}(\epsilon,\omega)$ is obtained using Dyson's equation:

\barray
&&\rho_{G_{el}}(\epsilon,\omega - \mu_{el}) = \rho_{\Sigma_{el} }(\omega- \mu_{el}) \times \nn\\
&&\frac{1}{(\omega-\Re e \ \Sigma_{el}(\omega- \mu_{el})-\epsilon)^2+\pi^2\left(\rho_{\Sigma_{el}}(\omega- \mu_{el})\right)^2}.\nn\\
\earray

The chemical potential $\mu_{el}$ is fixed using the number sum rule,

\beq
\int d\omega A_{el}(\omega - \mu_{el}) f(\omega - \mu_{el}) = \frac{n}{2},
\eeq
where $n$ is the density, and the onsite (i.e. local) spectral function
\beq
 A_{el}(\omega - \mu_{el})\equiv\int d\epsilon \ {\cal D}(\epsilon) \ \rho_{G_{el}}(\epsilon,\omega-\mu_{el}).
\label{Nomega}
\eeq

The self-energy due to the electron phonon interaction is given by
\barray
\rho_{\Sigma_{el,ph}} (\omega-\mu_{el}) &=& g^2\sum_{\pm} A_{el}(\omega\pm\omega_0 - \mu_{el})\times \nn\\
&&\left[f^{\mp}(\omega-\mu_{el}\pm\omega_0)+n_B(\omega_0)\right], \nn\\
\label{rhoSig}
\earray
\beq
\Re e \Sigma_{el,ph}(\omega) = \int d\nu  \ \frac{\rho_{\Sigma_{el,ph}}(\nu)}{\omega-\nu},
\eeq
where $f^{-}(\omega) \equiv f(\omega)$, $f^{+}(\omega) \equiv \bar{f}(\omega)$, and $n_B(\omega)\equiv \frac{1}{e^{\beta\omega}-1}$ is the Bose distribution function. Since $\omega_0\ll 1$, $A_{el}(\omega\pm\omega_0-\mu_{el}) \approx A_{el}(\omega-\mu_{el})$. In addition, for $T\gtrsim \omega_0$, $\sum_\pm f^{\mp}(\omega-\mu_{el}\pm\omega_0) \approx 1$ and $1+2n_B(\omega_0)\approx \frac{2T}{\omega_0}$. Therefore, for $T\gtrsim \omega_0$, to a very good approximation, 
\beq
\rho_{\Sigma_{el,ph}}(\omega-\mu_{el}) = \frac{\pi\lambda A_{el}(\omega-\mu_{el})}{2} \times T,
\label{rhoSigapprox}
\eeq
where $\lambda$, defined by $g^2\equiv\frac{\pi D \lambda \omega_0}{4}$, is a dimensionless measure of the electron-phonon coupling strength.

Putting everything together, the spectral function for the electron is given by
\barray
\rho_G(\epsilon,\omega-\mu_{el}) = \frac{\rho_\Sigma(\omega-\mu_{el})}{(R(\omega-\mu_{el})-\epsilon)^2+\pi^2(\rho_\Sigma(\omega-\mu_{el}))^2},\nn\\
\label{elspectral}
\earray
where $R(\omega-\mu_{el})\equiv \omega +\Delta\mu - \Re e \ \Sigma(\omega-\mu_{el})$, and  $\Delta\mu\equiv \mu-\mu_{el}$ tends to $0$ as $T\to\infty$,
\beq
 \Sigma(\omega) = \Sigma_{el}(\omega) + \Sigma_{el,ph}(\omega),
\eeq
and $\mu$ is determined by the condition
\beq
\int d\omega f(\omega - \mu_{el} ) A(\omega-\mu_{el})  = \frac{n}{2},
\label{sumrulemu}
\eeq
where
\beq
A(\omega-\mu_{el})  = \int d\epsilon  \ {\cal D}(\epsilon)  \rho_G(\epsilon,\omega-\mu_{el})
\eeq
is the LDOS.

Finally, the optical conductivity can be calculated from the spectral function via the formula:
\barray
\frac{\sigma(\omega)}{\sigma_{IRM}} = 2\pi  \int d\epsilon \phi(\epsilon)  \int && d\omega' \ \frac{\left[ f(\omega'-\mu_{el})-f(\omega+\omega'-\mu_{el})\right]}{\omega}  \nn\\
&& \times \rho_G(\epsilon,\omega'-\mu_{el}) \rho_G(\epsilon,\omega'+\omega-\mu_{el}),\nn\\
\;\;\; \;\;\label{optconddefn}
\earray
where $ \phi(\epsilon)= \Theta(1-\epsilon^2) \times (1-\epsilon^2)^{\frac{3}{2}}$ is the transport function. In the $\omega\to0$ limit, we recover Eqs. (5) and (6) in the MS for the dc conductivity.